\documentclass{PoS}

\usepackage{epsfig}

\title{Deep 20\,GHz observations of X-ray selected QSOs with the Compact Array Broadband
Backend (CABB)}

\ShortTitle{Deep 20\,GHz observations of X-ray selected QSOs with CABB} 

\author{\speaker{Elizabeth K. Mahony}\\
        Sydney Institute for Astronomy (SIfA), School of Physics, The University of Sydney, NSW 2006, Australia\\
	Australia Telescope National Facility, CSIRO, P.O. Box 76, Epping, NSW 1710, Australia \\
	E-mail: \email{emahony@physics.usyd.edu.au}}

\author{Elaine M. Sadler\\
        Sydney Institute for Astronomy (SIfA), School of Physics, The University of Sydney, NSW 2006, Australia \\}

\author{Scott M. Croom\\
        Sydney Institute for Astronomy (SIfA), School of Physics, The University of Sydney, NSW 2006, Australia \\}

\author{Ronald D. Ekers\\
	Australia Telescope National Facility, CSIRO, P.O. Box 76, Epping, NSW 1710, Australia \\}

\author{Ilana J. Feain\\
	Australia Telescope National Facility, CSIRO, P.O. Box 76, Epping, NSW 1710, Australia \\       }

\author{Tara Murphy\\
        Sydney Institute for Astronomy (SIfA), School of Physics, The University of Sydney, NSW 2006, Australia \\}

\abstract{The recently commissioned Compact Array Broadband Backend (CABB) on the Australia Telescope Compact Array (ATCA) provides 2\,GHz bandwidth in each frequency and polarisation, significantly increasing the sensitivity of the Array. This increased sensitivity allows for larger samples of sources to be targeted whilst also probing fainter radio luminosities. Using CABB, we have observed a large sample of objects at 20\,GHz to investigate the high--frequency radio luminosity distribution of X-ray selected QSOs at redshifts less than 1. Observing at high frequencies allows us to focus on the core emission of the AGN, hence recording the most recent activity. }

\FullConference{ISKAF2010 Science Meeting \\
                 June 10 -14 2010\\
                 Assen, the Netherlands}

\begin{document}

\section*{Introduction}

The recent upgrade to the Compact Array Broadband Backend (CABB) correlator on the Australia Telescope Compact Array (ATCA) has increased the bandwidth from 128\,MHz to 2\,GHz (for more details see paper by Emil Lenc in these proceedings). The increased sensitivity this provides allows us not only to go deeper, but to also observe large samples of sources in smaller amounts of time. To this effect, we have observed a sample of 1138 X-ray selected QSOs at 20\,GHz as a comprehensive study of the high-frequency radio luminosity distribution of QSOs.

QSOs are often classified into two broad categories; radio-loud and radio-quiet, but the underlying distribution of radio luminosity has long been debated in the literature. There are two opposing views - the first is that the distribution of radio-loudness is bimodal, i.e.\ there are distinct radio-loud and radio-quiet populations, with 10--20\% of QSOs being radio-loud\cite{kellermann}. The second view is that there is a broad, continuous distribution with no clear dividing line between radio-loud and radio-quiet QSOs\cite{cirasuolo}. Understanding this distribution can provide new insights into the high frequency radio properties of these sources while also allowing us to investigate the difference between X-ray selected and radio-selected QSO populations.

We are currently studying two samples of QSOs; one selected in X-rays using the {\it ROSAT} All Sky Survey (RASS), and the other selected in the radio from the Australia Telescope 20\,GHz (AT20G) survey\cite{murphy}. The two samples overlap in redshift, but only a small number of objects appear in both samples. We aim to compare the physical properties of radio and X-ray-selected QSOs in the redshift range $z<1$. As a first step, we have used the ATCA to make deep 20\,GHz observations of the $1138$ QSOs from the RASS X-ray sample which were not detected at the 40--50\,mJy flux limit of the AT20G survey. This will allow us to determine whether their radio luminosity distribution is continuous with that of radio-selected QSOs in the same redshift range, or is clumped at much lower radio luminosities (implying a bimodal distribution). 

By observing at 20\,GHz we pick up the central core component of the AGN and hence see the most recent activity. At lower frequencies we observe a higher fraction of emission from radio lobes; relics of past activity integrated over large timescales which therefore confuse our sample.

\section*{Sample Selection}

\begin{figure}[h]
\begin{minipage}{0.6\linewidth}
\centerline{\epsfig{file=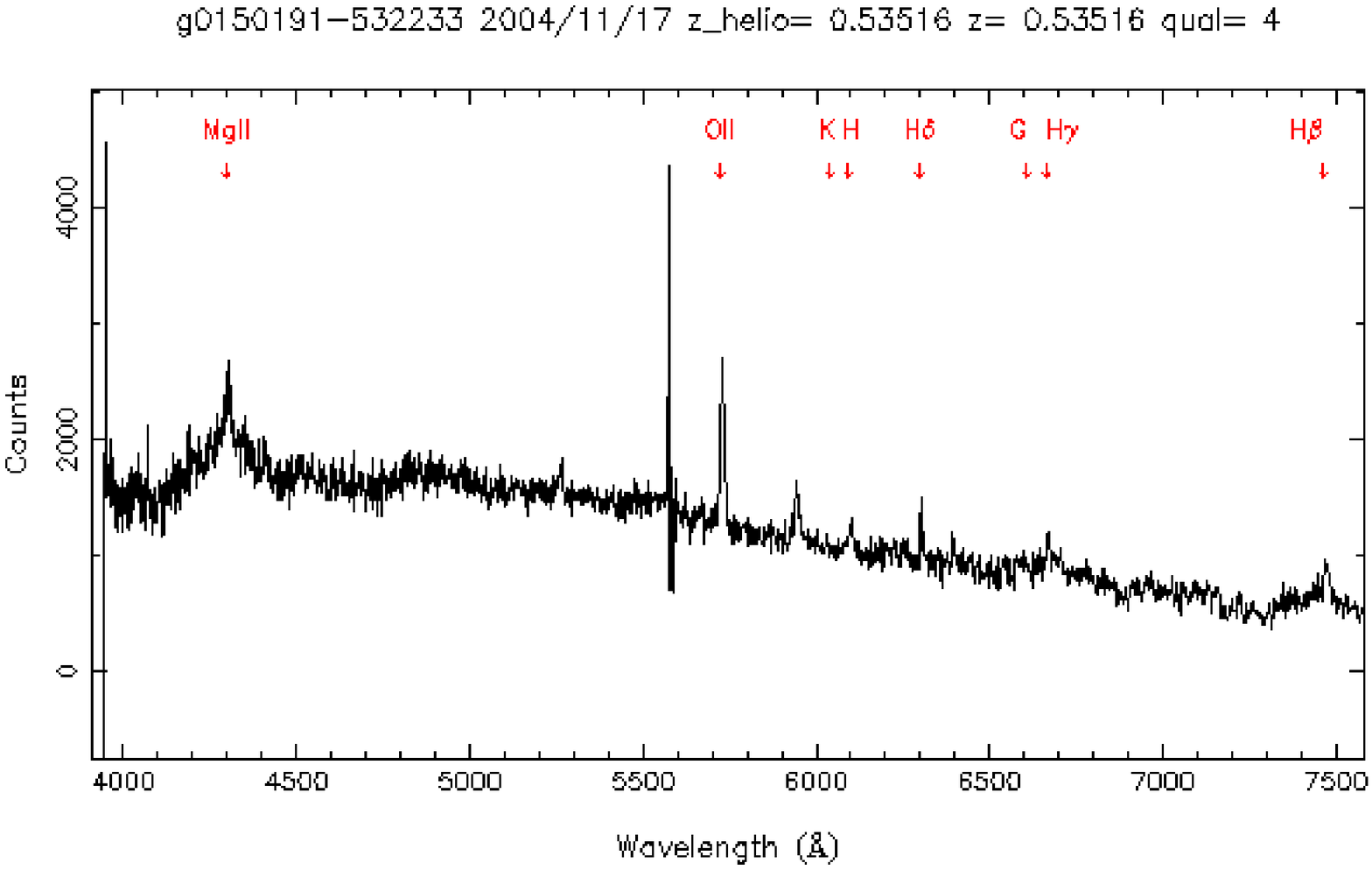, width=\linewidth}}
\end{minipage}
\begin{minipage}{0.4\linewidth}
\centerline{\epsfig{file=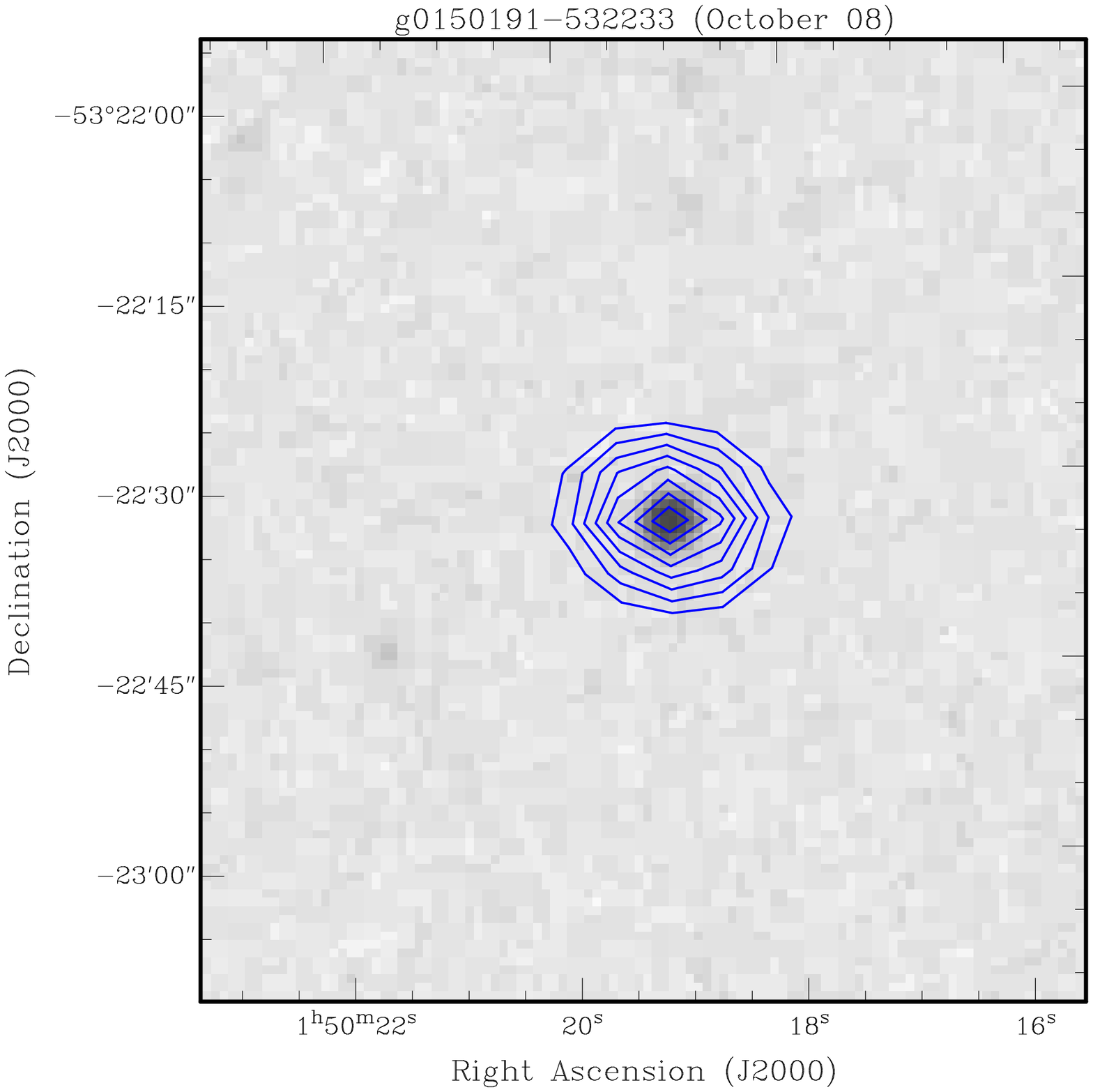, width=0.8\linewidth}}
\end{minipage}
\begin{minipage}{0.6\linewidth}
\centerline{\epsfig{file=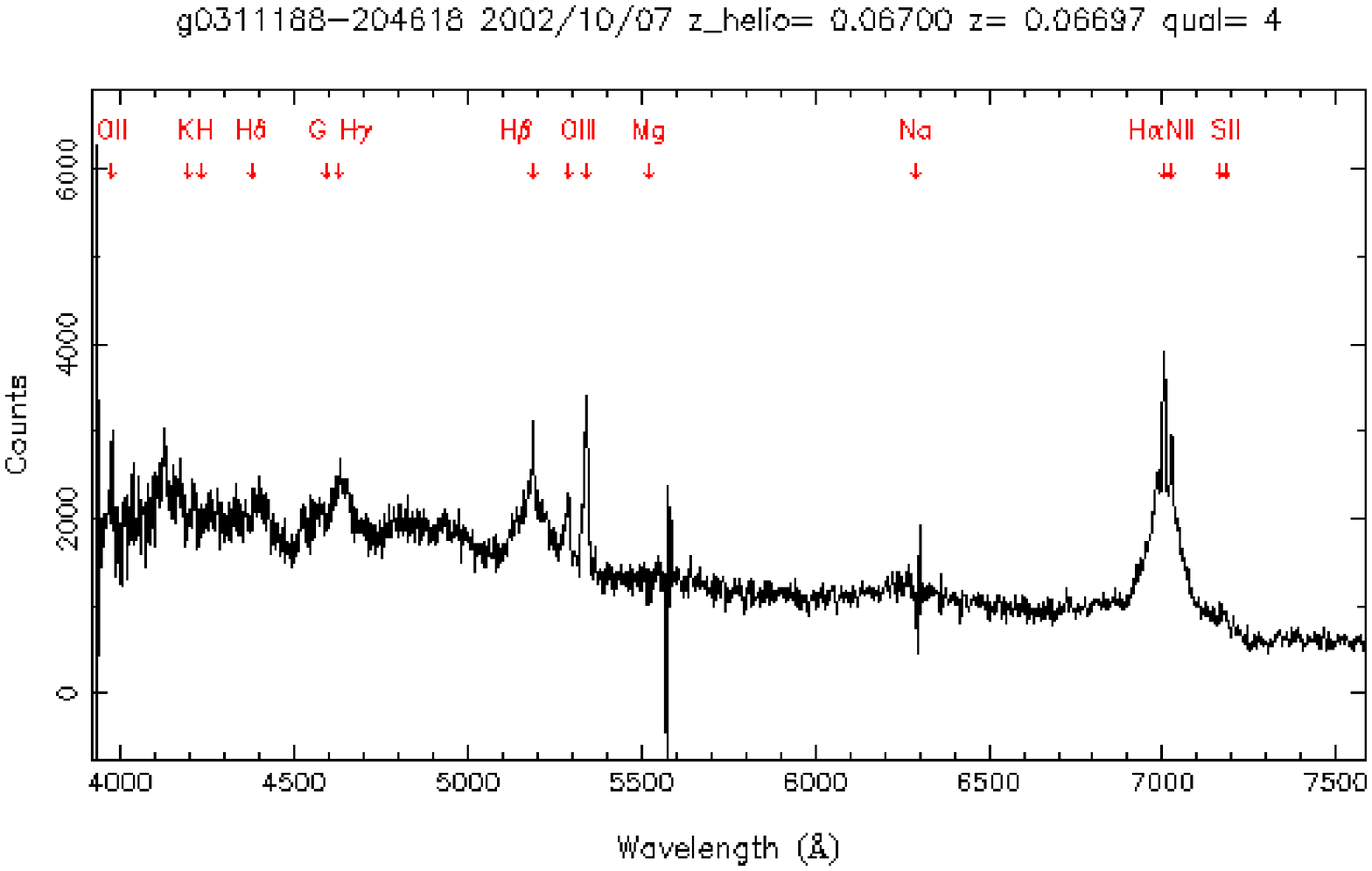, width=\linewidth}}
\end{minipage}
\begin{minipage}{0.4\linewidth}
\centerline{\epsfig{file=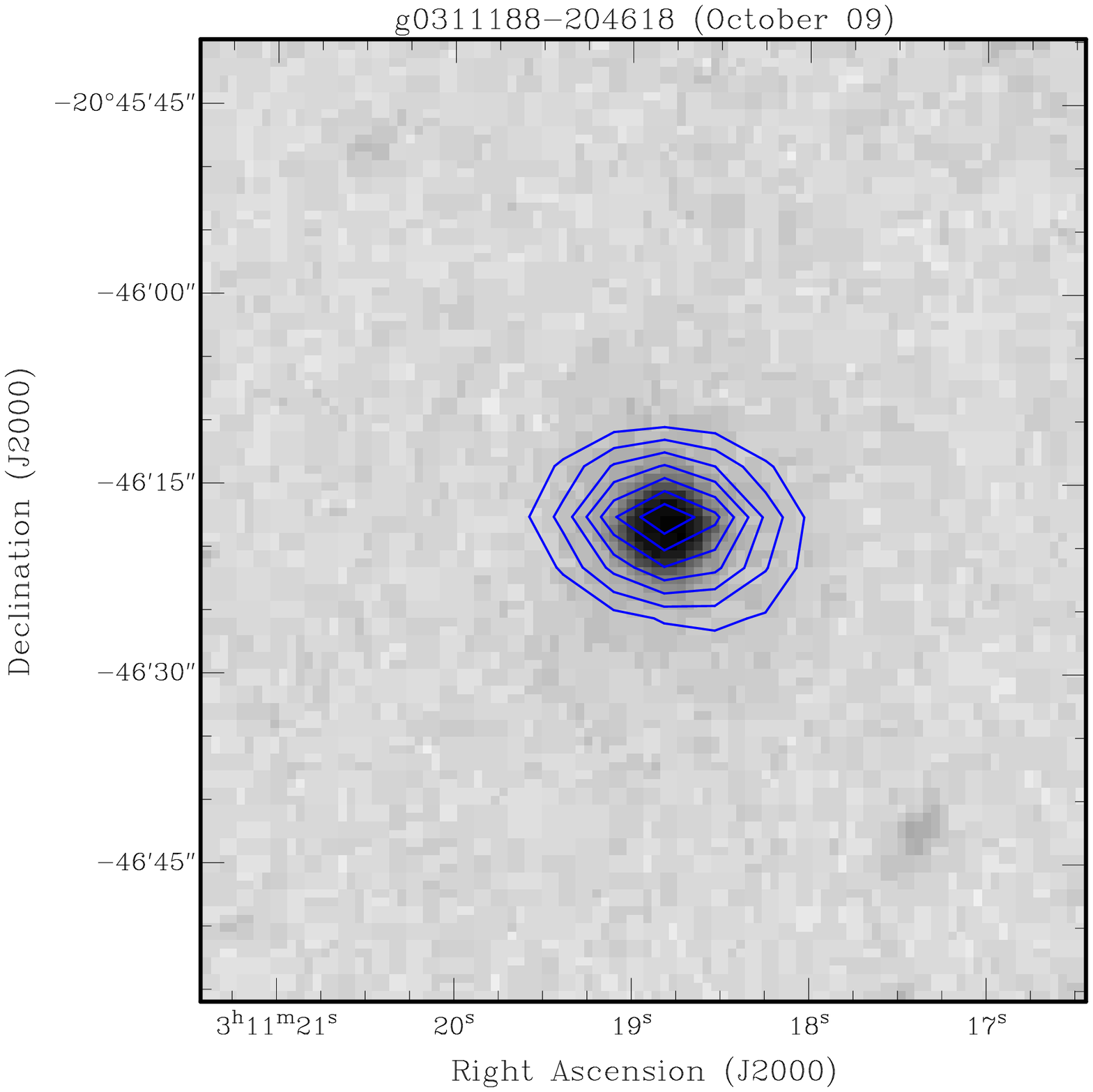, width=0.8\linewidth}}
\end{minipage}
\caption{Example RASS selected QSOs. The left image shows the 6dFGS spectrum. The arrows denote where spectral features would occur at that redshift, but not all of these are necessarily observed. On the right are the corresponding optical (B-band) images with 20\,GHz contours overlaid. The top source was observed in October 2008 (47.3\,mJy) with the old correlator and the bottom source was observed in October 2009 (1.64\,mJy) using CABB. In both images the contours represent 10\% intervals in flux, with the central contour corresponding to 90\%. }\label{example}
\end{figure}

Targets were seleced from the RASS--6dFGS catalogue\cite{mahony}; a catalogue of 3405 AGN selected from the ROSAT ALL Sky Survey (RASS) Bright Source Catalogue\cite{voges} that were observed as part of the 6dF Galaxy Survey (6dFGS)\cite{jones}. Sources were selected if the 6dFGS spectrum exhibited broad emission features indicative of a QSO or Type 1 AGN. We also set a redshift cutoff of z<1 to minimize any evolutionary effects. We then searched the Australia Telescope 20\,GHz (AT20G) survey for any known 20 GHz radio sources which were excluded from our target list. This leaves a final sample of 1138 X-ray selected QSOs at z<1. Example 6dFGS spectra and corresponding 20 GHz observations are shown in Figure \ref{example}. Since this is a large, low-redshift QSO sample which spans a wide range in optical luminosity, it can provide a definitive test of whether the $z<1$ QSO population is bimodal in its radio properties. Selecting sources that have 6dFGS spectroscopic information not only provides a uniform sample, but will also give us a wealth of extra information. Hence we can also study the optical spectral line properties, black hole masses, multiwavelength properties (X-ray -- optical --radio) and how these vary with redshift. 

\section*{Observations}

We observed 1138 X-ray selected QSOs at 20 GHz with the ATCA in a compact, hybrid configuration from 2008 -- 2010. The observations were carried out in a two-step process; all objects were observed for 2$\times$40s cuts and the objects not detected in that time were then reobserved for 2$\times$5min. These observations are summarised in Table 1. 
\begin{table}
\begin{center}
\caption{List of observations for this program. All the observations from 2008--2010 used the Hybrid 168m array configuration to obtain better u,v coverage. The October 2008 run was using the old correlator (128\,MHz bandwidth) and all other runs used CABB. \label{tab1}}
\begin{tabular}{lcccc}
\hline
{\bf Date of Observations} & \multicolumn{2}{c}{\bf No. Sources} & {\bf Time on Source} & {\bf Detection limit} \\
& {\bf Observed} & {\bf Detected} & & \\
\hline
October 2008 & 135 & 16 (12\%) & 2$\times$40s & 3\,mJy \\
April 2009 & 417 &  103 (25\%) & 2$\times$40s & 0.9\,mJy \\
October 2009 & 586 & 100 (17\%) & 2$\times$40s & 0.9\,mJy \\
& 377 & 58 (15\%) & 2$\times$5m & 0.5\,mJy \\
March 2010 & 122 & 21(17\%) & 2$\times$5m & 0.5\,mJy \\
\hline
\end{tabular}
\end{center}
\end{table}

\begin{figure}[h]
 \centerline{\epsfig{file=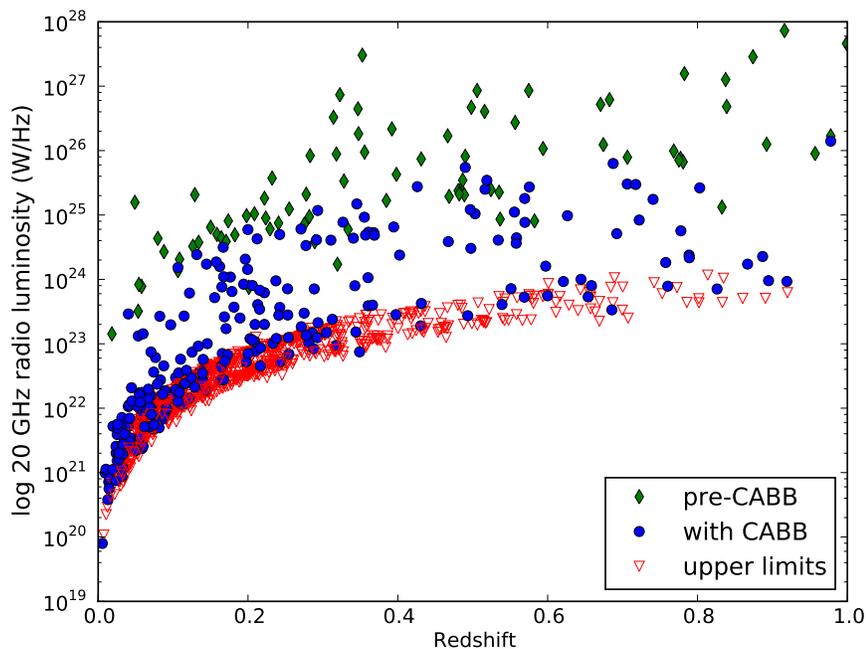, width=0.85\linewidth}}
\caption{20 GHz radio luminosity against redshift for our sample of X-ray selected QSOs. The green diamonds are sources that were observed using the old correlator on the Compact Array, including sources observed as part of the AT20G survey, while the blue circles show the sources that were detected at the 5 sigma threshold using CABB. The red triangles denote the upper limits for sources that weren't detected with CABB. }
\label{zrlum}
\end{figure}

\begin{figure}[h]
\begin{minipage}{0.5\linewidth}
 \centerline{\epsfig{file=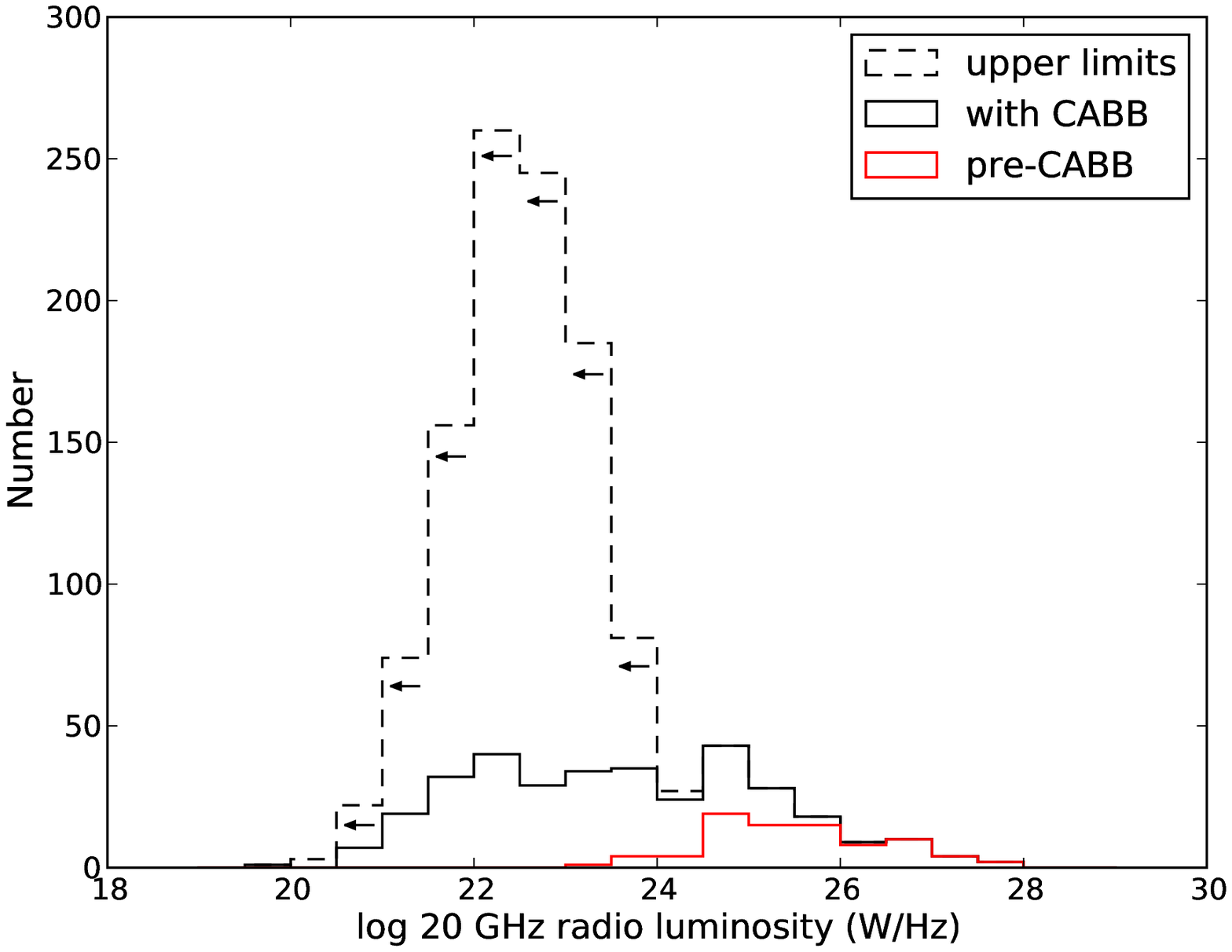, width=\linewidth}}
\end{minipage}
\begin{minipage}{0.5\linewidth}
 \centerline{\epsfig{file=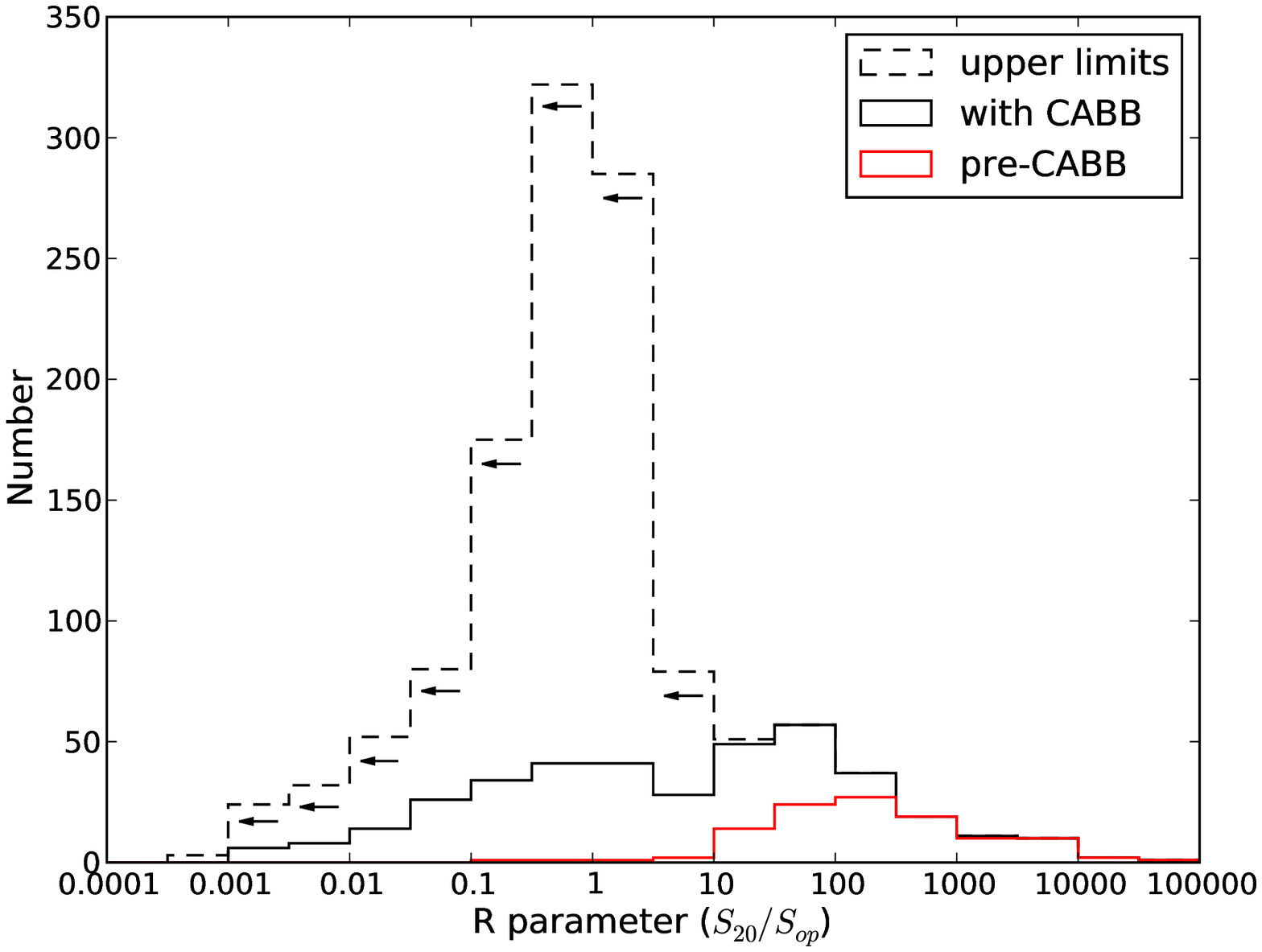, width=\linewidth}}
\end{minipage}
\caption{{\it(Left):} The radio luminosity distribution of our X-ray selected sample of QSOs at z<1. {\it(Right):} The `radio-loudness' distribution of this sample. The `radio-loudness' was determined using the Kellermann R-parameter which is defined as the ratio of the radio to optical fluxes. Radio-loud QSOs have R>10, while radio-quiet QSOs generally have R<1. Since all of our objects have optical spectroscopy and reliable redshifts, this allows us to compare both ways of determining whether a QSO is radio-loud and investigate any differences between them.}
\label{hist}
\end{figure}

Figure \ref{zrlum} shows preliminary results from these observations. The increased sensitivity provided by CABB is immediately obvious when comparing the 20\,GHz luminosities of sources detected using the old correlator (green diamonds) and those detected with CABB (blue circles). Figure \ref{hist} shows preliminary distributions of both the radio luminosity, and the R-parameter\cite{kellermann} which is often used as a measure of the `radio-loudness' of a QSO. This is defined as the ratio of the radio to optical flux; in this case the 20\,GHz radio flux divided by the optical B--band flux. In these figures, the red solid line indicates the distribution that was obtained using the old correlator on the ATCA whilst the solid black line shows the distributions obtained using CABB. The dashed line indicates the upper limits of the sources that were not detected. 

These figures, along with the detection limits noted in Table \ref{tab1} highlight the significant improvement that CABB has achieved.

\section*{Future Work}
The results presented here are very preliminary and the data reduction and analysis is ongoing\cite{mahony2}. However, the strict selection criteria and multiwavelength information provides a wealth of data allowing for a comprehensive study of these sources. In particular, work in progress includes:
\begin{itemize}
\item A statistical analysis of the data and resulting radio luminosity distributions.
\item Stacking experiments of the non-detected sources to study the average properties of radio-quiet X-ray QSOs. 
\item We also have data at 5 and 9 GHz for a subset of this sample, allowing us to investigate whether the radio luminosity distributions change with frequency.
\end{itemize}

\end{document}